\documentclass{emulateapj}

\usepackage[dvips]{epsfig,color}
\usepackage{gensymb}

\newcommand{\eg}{{\it e.g.,}}
\newcommand{\ie}{{\it i.e.,}}
\newcommand{\etal}{{\it et al.}}

\newcommand{\ignore}[1]{\relax}
\newcommand{\ba}{\ensuremath{\frac{b}{a}}}
\newcommand{\ca}{\ensuremath{\frac{c}{a}}}
\newcommand{\mur}{\ensuremath{\mu_r}}
\newcommand{\murz}{\ensuremath{\mu_{r,0}}}
\newcommand{\mui}{\ensuremath{\mu_i}}

\newcommand{\mut}{\ensuremath{\mu_t}}

\newcommand{\muic}{\ensuremath{\mu_{i,c}}}
\newcommand{\murmax}{\ensuremath{\mu_{r,{\rm max}}}}
\newcommand{\tc}{\ensuremath{\: t_{\rm cross}}}
\newcommand{\qz}{\ensuremath{Q_0}}

\shorttitle{}
\shortauthors{Barnes \etal}

\slugcomment{Accepted for publication in ApJ}

\begin{document}

\title{The Radial Orbit Instability in Collisionless N-Body
Simulations}
\author{Eric I. Barnes}
\author{Paul A. Lanzel}
\affil{Department of Physics, University of Wisconsin --- La Crosse,
La Crosse, WI 54601}
\email{barnes.eric@uwlax.edu}
\email{lanzel.paul@students.uwlax.edu}
\author{Liliya L. R. Williams}
\affil{Department of Astronomy, University of Minnesota,
Minneapolis, MN 55455}
\email{llrw@astro.umn.edu}

\begin{abstract}

Using a suite of self-gravitating, collisionless $N$-body models, we
systematically explore a parameter space relevant to the onset and
behavior of the radial orbit instability (ROI), whose strength is
measured by the systemic axis ratios of the models.  We show that a
combination of two initial conditions, namely the velocity anisotropy
and the virial ratio, determines whether a system will undergo ROI and
exactly how triaxial the system will become. A third initial
condition, the radial shape of the density profile, plays a smaller,
but noticeable role.  Regarding the dynamical development of the ROI,
the instability a) begins after systems collapse to their most compact
configuration and b) evolves fastest when a majority of the particles
have radially anisotropic orbits while there is a lack of
centrally-concentrated isotropic orbits. We argue that this is further
evidence that self-reinforcing torques are the key to the onset of the
ROI.  Our findings support the idea that a separate orbit instability
plays a role in halting the ROI.

\end{abstract}


\section{Introduction}\label{intro}

$N$-body simulations of self-gravitating, collisionless systems of
massive particles have become a standard tool used to investigate the
behavior of galactic systems.  Such particle aggregations can
represent the visible, stellar components of galaxies as well as the
dark matter that is hypothesized to surround individual galaxies
(halos).  Since the dark matter halos of galaxies are thought to
contain the majority of the mass in a galactic system, the simplest
approximation to modeling a real galaxy is to ignore the stars and
focus on the dark matter.

This approach has led to cosmological simulations that produce cosmic
webs of dark matter, where hierarchical merging creates sub-galactic,
galactic, and cluster sized systems.  More sophisticated models
\citep[like the Millenium Simulation,][]{setal05} also include gas and
stars, but the dark matter remains the dominant gravitational
component.  Simulations like these have found a number of interesting
properties of dark matter systems.  The density profiles of dark
matter halos tend to have shapes described by a simple relation
\citep[\eg][]{n04}.  Generically, density increases with decreasing
radius, but the increase is larger near the edges of systems.  The
overall shape of the density profiles appears to be a robust outcome
of dark matter halo evolution \citep{lh03,letal06,gc07}.  However, the
inclusion of gaseous baryonic material can impact the innermost
regions of dark matter halos \citep{ketal06}, possibly playing an
important role in halo evolution.

Our overall goal is to better understand the physics involved in the
evolution of dark matter halos.  Such understanding should explain,
for example, the shapes of halo density profiles and why the
coarse-grained phase-space density is single power-law in radius
\citep{tn01,b07a}.  In this work, we focus on models that involve, to
some degree, the well-discussed radial orbit instability (ROI)
\citep[for example,][]
{ps81,va82,ma85,pp87,k91,s91,cnm96,ts99,hjs99,b05,mwh06,ba06,betal08}.
To this end, we have constructed a set of $N$-body models with a
variety of defining characteristics to represent dark matter halos.
These models have been evolved in time, and the evolutions have been
analyzed to find common trends in behavior.

The ROI can be briefly summarized as follows.  Initally spherical
$N$-body systems made up of particles that have predominantly radial
motions do not remain spherical.  In general, the overall shapes of
these systems become prolate spheroidal or triaxial \citep[but on time
scales of order the relaxation time, they can become more spherical
again, see][]{ts99}.  The instability appears in models with and
without Hubble expansion.  As this instability can drive changes in
the density and velocity distributions in dark matter systems, we are
interested in furthering the description and understanding of its
mechanisms.

We will present relationships between initial conditions of $N$-body
models and the behavior of the ROI.  In particular, we show that
neither global measures of velocity anisotropy nor virial ratio alone
can be used to predict the onset of the ROI.  Instead, we find that
combining these measures provides a determination of the end result of
the ROI, specifically the overall shape of the system.  We also argue
that the results of our simulations support the idea that another
orbit instability, detailed by \citet{aetal07}, plays an important
role in halting the ROI.  However, we point out that our simulations
cannot speak to any possible role gas could play in stopping the ROI.

As a first step in gathering information about the physics of
collisionless, self-gravitating systems, we investigate models that do
not include Hubble expansion.  As we will discuss, these
non-cosmological systems contain complexities that would be compounded
by the inclusion of initial radial expansion.  Section~\ref{methods}
begins with an explanation of our model creation and evolution
methods.  Our analysis routines and testing criteria are also
described in this section.  The crux of this work begins with
\S~\ref{globroi}, a discussion of the global behavior of the ROI in
our models.  Section~\ref{roidev} deals with our observations of the
onset and early development of the ROI.  Executive summaries of these
topics can be found in Sections~\ref{globsynth}, \ref{anisopic}, and
\ref{isopic}.  A synopsis of this work and our conclusions are
presented in \S~\ref{summary}.

\section{Methods \& Testing}\label{methods}

\subsection{Initial Conditions}\label{inits}

\begin{figure}
\plotone{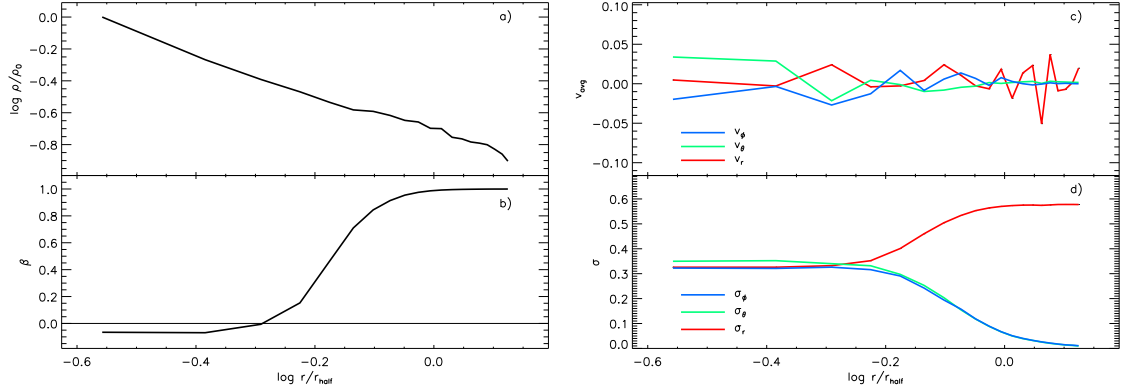}
\figcaption{We illustrate important initial condition information for
an anisotropic system with a cuspy density profile ($\rho \propto
r^{-1}$) and $\qz=0.5$.  The log-log density profile of the system is
shown in panel a.  Panel b shows the anisotropy profile.  The average
spherical component velocity profiles and velocity dispersion profiles
are illustrated in panels c and d, respectively.
\label{initfig}}
\end{figure}

We create $N$-body systems with three different initial density
profiles; $\rho=\rho_0$ (constant), $\rho \propto r^{-1}$ (cuspy), and
$\rho \propto e^{-r^2}$ (Gaussian).  Particles are distributed
randomly within a sphere of radius $R=1$ according to the density
distribution so that the total mass $M=1$.  Both $R$ and $M$ are
discussed in dimensionless code units where $G=1$; fixing a physical
value of $M$ and $R$ would allow us to also transform any other
quantities to physical units.  The cuspy and Gaussian profiles involve
scalelengths; the density at the scalelength is half its central value
for the cuspy profile and $e^{-1}$ times its central value for the
Gaussian profile.  The cuspy scalelength is $10^{-3}$ and the Gaussian
scalelength is $1/\sqrt{5}$.  The density profile of a cuspy system is
shown in Figure~\ref{initfig}a.  For most of our simulations, the
particles have equal masses, but we have also created models with
particles of different mass (see \S~\ref{test}).  For these multi-mass
models, the fractions of particles with each mass is determined so
that the total mass of the system remains $M=1$.

With the initial positions of the particles fixed, we calculate the
potential energy of a system and adopt a value for the initial
virial ratio $\qz \equiv T/|W|$ to calculate a scale speed that is
used to assign individual particle velocities.  The virial ratio
changes during the evolution of the systems, and we denote its value
at times after $t=0$ simply as $Q$.  Velocities are assigned to
particles based on an assumed radial velocity anisotropy profile,
where $\beta=1-\sigma^2_{\rm tan}/2\sigma^2_{\rm rad}$ is the velocity
anisotropy and $\sigma$ denotes velocity dispersion.  We utilize the
anisotropy profile discussed in \citet{b07a},
\begin{equation}
\beta(r)=\frac{1}{2}(\beta_{\rm high}-\beta_{\rm
low})[1+\tanh{(m \log{r/r_a})}]+\beta_{\rm low},
\end{equation}
where $\beta_{\rm low}$ is the anisotropy value at $r=0$, $\beta_{\rm
high}$ is the value at large $r$, $m$ controls the transition between
these values, and $r_a$ is the anisotropy radius, at which
$\beta=0.5$.  The flexibility of this relation allows us to easily
create completely isotropic models ($\beta=0$), completely radially
anisotropic models ($\beta=1$), or systems with intermediate levels of
radial anisotropy.  With fixed values of $\beta_{\rm low}$ and
$\beta_{\rm high}$, the amount of anisotropy is controlled mainly by
the value of $r_a$, but the value of the slope $m$ also contributes.
For the work discussed here, we have fixed the slope value $m=7$ to
provide a relatively rapid change between $\beta_{\rm low}$ and
$\beta_{\rm high}$.  Note that a given anisotropy profile will provide
different amounts of mass with radially anisotropic velocities
depending on the density profile.  An example $\beta(r)$ with
$\beta_{\rm low}=0$, $\beta_{\rm high}=1$, and $r_a=0.5$ is plotted in
Figure~\ref{initfig}b.

For a chosen anisotropy profile, we proceed to assign individual
velocities to particles.  A particle's speed can be taken from a
distribution about the scale speed.  In isotropic regions, particles
have velocity directions that uniformly cover $4\pi$ steradians.
Radial anisotropy is guaranteed by limiting the velocity vector to lie
within cones centered on a particle with axes along the radial
direction (one cone opens towards the center, the other opens
outward).  The opening angle of the cone correlates to the desired
value of $\beta$.  For small $\beta$, the angle is close to $\pi$;
when $\beta \approx 1$, the angle is nearly 0.  The velocity vector
for a specific particle is chosen randomly within the allowed range of
directions.  Since the vector has definite limits placed on its
direction, we refer to this as the `hard-edge' setup.  Panels c and d
of Figure~\ref{initfig} show the average component velocities and
dispersions, respectively, for a model with the same anisotropy
parameters as above and $\qz=0.5$.  The average component velocities
for all shells in all models are initially zero.

Of course, this is not the only route to initially anisotropic
systems.  We have also created models with initial conditions that
contain combinations of isotropic and radially anisotropic orbits at
all radii, in contrast to the `hard-edge' setup.  The evolutions of
these `soft-edge' models is very similar to those using the
`hard-edge' setup.  All initially anisotropic models subsequently
discussed use the `hard-edge' setup.

\subsection{Evolution \& Analysis}\label{evo}

We have used the direct $N$-body integration code NBODY2 and the
treecode Gadget-2 to evolve our ensembles of particles.  Readers
interested in the details of NBODY2 and Gadget-2 algorithms are
encouraged to examine \citet{a01} and \citet{s05}, respectively.  Any
two particles $i$ and $j$ interact through softened forces of the
form,
\begin{equation}
\vec{F}_{ij}=-G \frac{m_i
m_j}{(r_{ij}^2 + \epsilon^2)^{3/2}}(\vec{r}_i-\vec{r}_j),
\end{equation}
where $r_{ij}$ is the distance between the particles and $\epsilon$ is
the softening length.  NBODY2 allows particle-particle merging, but we
do not implement that option in our models.  The time interval in
these models is the crossing time determined by the initial density
profile.  A typical run evolves the system for approximately 20 inital
crossing times $\tc$.  The actual length of an evolution depends on
the strength of the collapse, with stronger collapses running for
fewer crossing times as close particle encounters decrease timestep
sizes.  However, all simulations have been run long enough for the
system to come to virial equilibrium, and most have reached mechanical
equilibrium throughout.  Some models have an outer region containing
$\approx$ 5\% of the total mass that continues to expand throughout
the simulations.  NBODY2 includes a routine for removing particles
that become unbound from the system, however Gadget-2 does not.  The
direct nature of NBODY2 limits the particle numbers that we can
reasonably investigate (in this work $N=10^4$).  As a treecode,
Gadget-2 allows us to evolve models with an order of magnitude more
particles ($N=10^5$) in roughly the same amount of time.  Aside from
the ability to remove escapers with NBODY2, the biggest difference
between the two codes is that energy conservation is much better with
NBODY2 (but adequate with Gadget-2).

Output from both codes occurs at fixed intervals determined by the
initial crossing time of the system \tc.  We have modified the NBODY2
ouput process somewhat to keep track of different variables than
originally intended.  We create ``snapshots'' of the system at the
output intervals; numbers of particles, postions, velocities,
energies, numbers and strengths of collsions, etc.  Gadget-2 output
files containing particle positions and velocities are
straightforwardly generated.  Once a model has completed its
evolution, the snapshots are analyzed to determine the evolution of
quantities such as density, velocity dispersion and anisotropy, and
axis ratios.

Our analysis tools generally come in two groups, spherical and
ellipsoidal, but we utilize only spherical tools in this work.
Our analyses place particles in shells that each contain a fraction of
the total system mass.  For NBODY2 models, each shell contains 5\% of
the bound mass, while Gadget-2 shells contain 1\% of the total mass.
These shells are used to determine average densities, average
velocites and dispersions, etc.  that give us radial profiles of these
quantities at any point during the run.  Additionally, two sets of
systemic axis ratios are determined.  These axis ratios relate the
length of the longest axis of a system $a$ to the lengths of the
intermediate $b$ and short $c$ axes.  One pair of (\ba,\ca) values is
created using positions of particles that make up the innermost 95\%
of the mass in a halo.  The outer 5\% of the mass has been discounted
as it can often reach distances greater than 10 half-mass radii.
Particles at such large distances can allow small amounts of mass to
severely distort the axis ratios from values appropriate for the more
centrally concentrated mass of the system.  The other axis ratio pair
gives the shape of the innermost 80\% of the mass.  Comparing the
behaviors of these two sets of axis ratios gives us leverage to
understand which part of the halo (inner or outer) is dominating the
overall shape.  We note that since Gadget-2 does not remove unbound
particles, 95\% mass axis ratios from Gadget-evolved systems tend to be
rounder than those derived from NBODY2 models.

We also track velocity quantities throughout the evolutions.  Average
spherical component velocities ($v_r$, $v_{\phi}$, $v_{\theta}$) and
velocity dispersions ($\sigma_r$, $\sigma_{\phi}$, $\sigma_{\theta}$)
are calculated for the particles in a given shell.  With the
dispersions, we calculate the anisotropy parameter $\beta$ for each
shell (effectively creating a radial profile).  We divide the
anisotropy categories as follows: if the anisotropy of a shell $\beta
> 0.5$, it is radially anisotropic; isotropic shells have $-1.0 <
\beta < 0.5$, and $\beta < -1.0$ for a shell to be tangentially
anisotropic.  We have chosen these boundary values since $\beta =0.5$
occurs when the radial dispersion is double that for either of the
tangential directions ($\sigma_r=2\sigma_{\phi}=2\sigma_{\theta}$).
Similarly, $\beta=-1.0$ when each of the tangential dispersions are
twice the radial dispersion
($\sigma_{\phi}=\sigma_{\theta}=2\sigma_r$).

We keep track of the fractions of mass with isotropic (\mui), radially
anisotropic (\mur), and tangentially anisotropic (\mut) orbits by
adding the particle numbers of shells with appropriate $\beta$ values.
We also calculate the fraction of centrally-concentrated mass that is
isotropic \muic, where a centrally-concentrated shell must be part of
the innermost 50\% of the mass of a system.  The mass fractions are
related through $\mur + \mut + \mui = 1$, and while \mui\ can be as
large as 1, \muic\ has a maximum value of 0.50.

\subsection{Testing}\label{test}

We have tested the robustness of our NBODY2 results by varying two
important parameters, softening length and number of particles.  The
NBODY2 softening length has been varied by two orders of magnitude
($\epsilon=5\times10^{-2}$ to $\epsilon=5\times10^{-4}$).  Pairs of
models with identical initial conditions and parameter sets, save
softening length, show that softenings in this range lead to
essentially identical evolutions.  Gadget-2 simulations use
$\epsilon=1\times10^{-4}$.  We note that these values of $\epsilon$
are approximately the range suggested by \citet{p03}.

\citet{bak02} and \citet{ba06} argue that a conservative estimate of
$N\approx 10^5$ is required to accurately follow non-spherical
collapses.  To further test the impact of resolution on the quantities
relevant for this work, we have also re-evolved 3 versions of
$N=10^4$ models using $N=5\times10^4$ with NBODY2.  These specific
models were chosen to represent the range of initial density profiles
(constant to Gaussian), velocity anisotropy (isotropic to radially
anisotropic), and kinetic temperatures (very cold to warm).  The
$N=5\times 10^4$ models behave very similarly to the $N=10^4$ models.
We also ran a single $N=10^5$ run with NBODY2 as an extreme test.
The most important quantities for this work, overall strength of the
ROI in terms of axis ratios, rate of ROI onset, and velocity dispersion
evolution, do not appear to change dramatically with increased $N$.
Most importantly, the Gadget-2 models with $N=10^5$ do not show any
coordinated or significant deviations in behavior from their NBODY2
counterparts.

The softening value plays a key role in the degree of collisionality
of the evolutions.  If $\epsilon$ is too small, close encounters
between particles give rise to unwanted two-body effects.  When
$\epsilon$ is too large, the fundamental interaction between particles
ceases to be Newtonian gravity.  We have tested the collisionality of
our halos in two ways, using NBODY2.  First, we have tracked the
changes in particle velocities $\Delta \vec{v}$ throughout our
simulations.  Strong two-body collisions lead to values of $|\Delta
\vec{v}|/v \approx 1$.  We find that only a small fraction of the
particles in our models undergo such strong collisions.  Second,
because collisional effects would drive mass segregation on the
two-body relaxation timescale, we have evolved multi-mass models where
particles are divided into two groups with different masses (in our
tests, the mass ratio is 3).  Two multi-mass models with softenings
$\epsilon=5\times10^{-2}$ and $\epsilon=5\times10^{-4}$, respectively,
have been evolved for approximately one two-body relaxation time
($\approx 140 \: \tc$).  The evolutions show virtually identical
behaviors; in each run, the half-mass radius of the heavier particles
quickly reaches a value that remains virtually constant up to one
relaxation time.  The lack of mass segregation indicates that our
simulations within this softening range are collisionless for at least
100 crossing times.

These tests demonstrate that the essential physics of collisionless,
self-gravitating collapses is present in our simulations.  Of course,
larger particle numbers would allow higher resolution, but our goal is
not to push the boundaries of numerical simulation.  These
straightforward simulations allow us to focus on understanding the
physics of the radial orbit instability by investigating a large
variety of initial conditions.

\section{Global Behavior of the Radial Orbit
Instability}\label{globroi}

The most common criteria for determining whether or not the radial
orbit instability has occurred involves the overall shape of the
system.  If a system remains more-or-less spherical throughout its
evolution, the ROI has not developed.  When a system changes from an
initial spherical shape to a final spheroidal or triaxial shape, the
ROI has occurred.  The weakness of previously defined quantitative
boundaries that separate ROI/non-ROI models has lead us to ask if such
boundaries exist and, if so, how they may be better described.  

We are looking for quantities that distinguish between initial
conditions that lead to ROI and non-ROI situations.  It has been
suggested that the global quantity $A \equiv 2T_r/T_{\rm tan} =
2<\sigma^2_r>/<\sigma^2_{\rm tan}>$ is a good ROI diagnostic
\citep[\eg][]{ps81,bgh86,ma85,betal08}.  From analytical
considerations, \citet{fp84} and \citet{ps81} suggest that an initial
value of $A_0 \ga 1.7$ will ensure that a system will undergo the ROI.
However, the value of $A$ associated with the ROI has been found to
range from $\approx 1.4$ to $\approx 3.0$, and \citet{u93} and
\citet{hetal96} discuss situations in which the ROI occurs despite $A$
having relatively small values.  As \citet{bgh86} speculate, this
range is at least partially caused by different initial distributions
of mass and velocity.  Also, \citet{pp87} point out that this quantity
does not have a critical value that separates stable and unstable
systems, but that the growth rate of the instability may become small
enough as $A$ decreases that systems are practically stable.  We
investigate whether or not other global measures can provide a clearer
demarcation between stable and unstable systems.  In particular, we
use the initial virial ratio $\qz$ and the initial fraction of mass
with radially anisotropic velocities \murz.  As we discuss below, $A$
and \mur\ are different ways of measuring roughly the same
characteristic of a system, however we focus on \mur\ as it has
well-defined limits and a simple interpretation.

To clarify the notation that we use in this paper, the variables $Q$,
$A$, \mur, \mui, and \mut\ refer to the value of those quantities at
arbitrary times during an evolution.  We are often interested in the
initial values of those quantities, which we will denote with a
subscript `0'; for example, the initial value of the virial ratio is
\qz. 

As with previous investigations, we find that two seemingly different
classes of initial conditions lead to the ROI; systems with initial
velocity anisotropy and dynamically cold, but isotropic systems.  The
subsequent discussions of these two classes focus on their common
aspects as a way to understand the basic physics of the ROI.  We will
discuss the development of the ROI for the two cases in
S~\ref{roidev}, but for the moment we focus on whether or not the two
cases lead to different global behaviors.

\subsection{Initially Anisotropic Systems}\label{anisomods}

Models that have enough radial anisotropy undergo the ROI; \ie the
shape of the model becomes non-spherical.  We quantify the presence
and strength of the ROI in terms of the values of the systemic
axis ratios.  Our goal is to find a link between the amount of radially
anisotropic mass \murz\ initially present in the models and the
strength of the instability.  For each of the initial density profiles
discussed in \S~\ref{methods}, we follow models with $\qz=1.0$ (hot),
0.5 (warm), 0.2 (warm), and 0.1 (cold).  Twelve models with $0.1 \le
r_a \le 1.2$ are created for each $\qz$, covering a spectrum of \murz\
values from nearly isotropic to completely radially anisotropic.  We
have found that the radially anisotropic mass fraction has a
monotonic, one-to-one relationship to the global anisotropy parameter
$A$.  The exact correspondence depends on the particular density
profile of the model, but it is not surprising that nearly isotropic
models with small \mur\ have $A\approx 1$, while models with $\mur
\approx 1$ have $A \gg 1$.  Thus, the span of \murz\ values for our
models implies a corresponding range of coverage in terms of $A$.

We use the innermost 80\%-mass axis ratios found in each simulation as
measures of the strength of the instability.  Specifically, the time
at which \ca\ is minimum is found and then the corresponding \ba\
value is saved.  While it is common for \ba\ and \ca\ to have their
minimum values at the same time, this is not always the case.  In
general, the time of minimum \ca\ is approximately the virialization
time (when $Q$ begins to maintain a steady value of 0.5).

\begin{figure}
\plotone{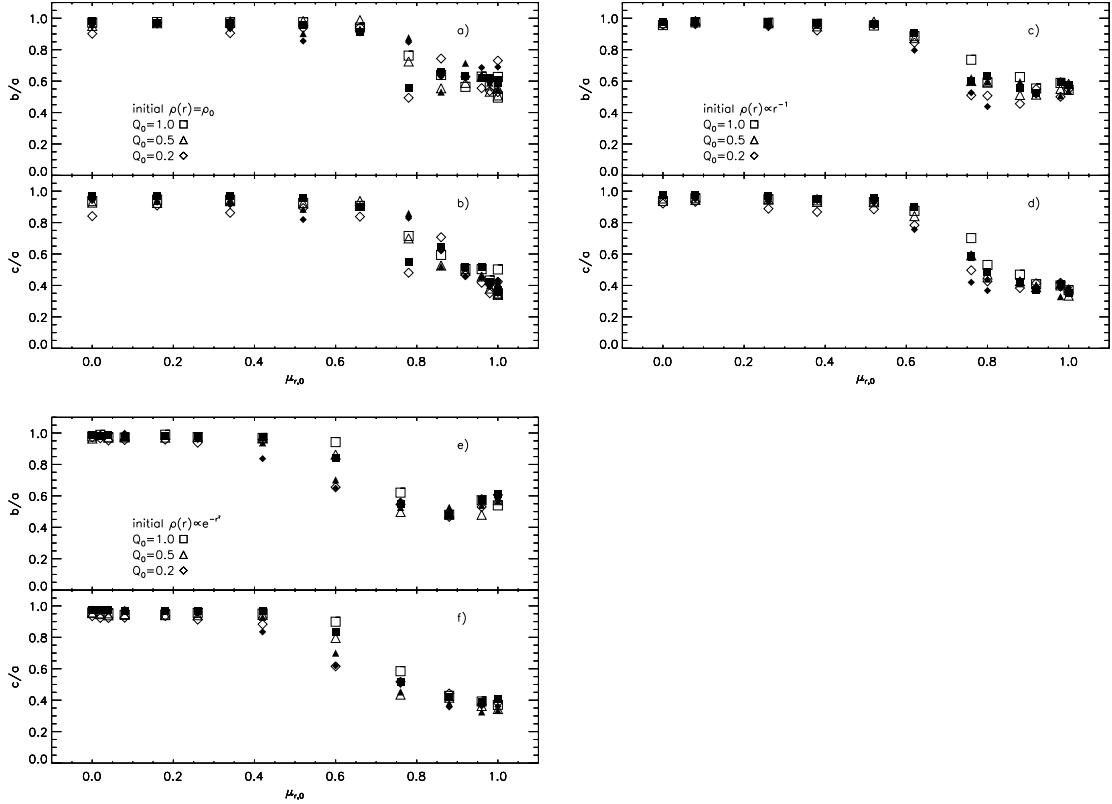}
\figcaption{Minimum axis ratios for the innermost 80\% of mass in
models that have initial radial anisotropy and virial ratios $\qz \ge
0.2$.  In each panel, models with $\qz=1.0$ are denoted by square
symbols, $\qz=0.5$ models are marked with triangles, and systems with
$\qz=0.2$ are represented by diamonds.  The empty symbols are derived
from the NBODY2 simulations with $N=10^4$ particles, while the filled
symbols represent Gadget-2 simulations with $N=10^5$ particles.  The
similar patterns followed by initially constant density models are
shown in panels a) and b).  Panels c) and d) show the similarities
between axis ratio behavior in initially cuspy models.  The same basic
patterns also appear for initially Gaussian density models, panels e)
and f). As the initial amount of radial velocity anisotropy increases,
models become non-spherical.
\label{hiqglob}}
\end{figure}

Hot and warm models ($\qz \ge 0.2$) generally have the same behavior,
independent of initial density profile (Figure~\ref{hiqglob}).  In
this figure, the open symbols are derived from $N=10^4$ NBODY2
simulations and the filled symbols result from $N=10^5$ Gadget-2
simulations.  There is no significant difference between the two.  For
$\murz \la 0.60$, the systems remain nearly spherical.  Higher amounts
of initial radially anisotropic orbits lead to non-spherical systems.
We refer to the \murz\ values that separate spherical and
non-spherical models as threshold values.  For models with $0.60 \la
\murz \la 0.80$, the systems take on prolate spheroidal shapes when
the instability has maximum strength.  Systems that are nearly
completely dominated by radially anisotropic orbits tend to take on
fully triaxial shapes at the peak of the instability.  The \ca\ values
of these heavily radial models tend to decrease as \murz\ increases,
but \ba\ does not, leading to triaxiality.  Overall, there is a slight
trend for our models, at all \murz, to become less spherical as they
become colder.

\begin{figure}
\plotone{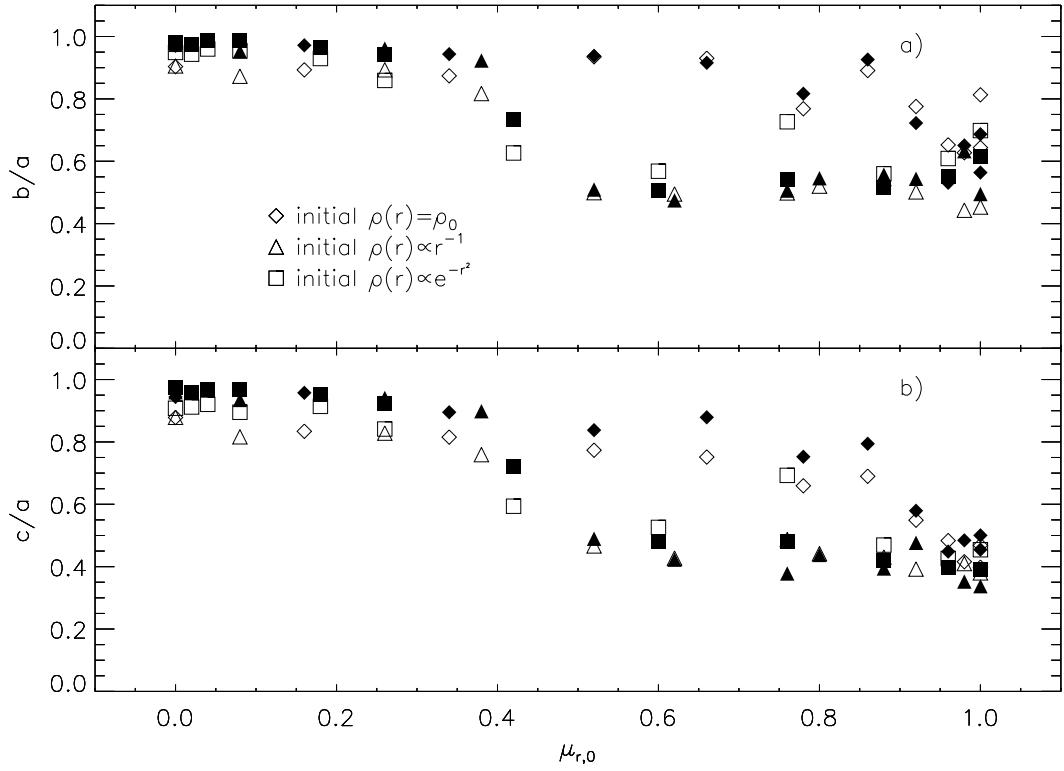}
\figcaption{Minimum 80\%-mass axis ratios for cold $\qz=0.1$, initially
anisotropic systems.  As in Figure~\ref{hiqglob}, the open and closed
symbols refer to evolved with NBODY2 and Gadget-2, respectively.  The
cuspy and Gaussian models (denoted by triangles and squares,
respectively) follow the same basic trend for both \ba\ and \ca.
Models with initially constant density (represented by diamonds) do
not show much variation with \mur, but these systems represent the
remnants of models with significant mass loss.
\label{q0.1glob}}
\end{figure}

From Figure~\ref{q0.1glob}, we see that the coldest initially
anisotropic models with $\qz=0.1$ depart from the unified behavior of
the warm models.  As in Figure~\ref{hiqglob}, the open and filled
symbols are derived from NBODY2 and Gadget-2, respectively.  The
initially constant density models barely remain spherical even when
mostly isotropic, and there is no sudden transformation to
substantially non-spherical shapes.  Overall, \ba\ values do not
change much with \murz, and \ca\ decreases only slowly as \murz\
increases.  Models with $\murz \ga 0.60$ become triaxial when the ROI
is at peak strength.  The cuspy and Gaussian models continue to show
the previously noted transition, but the details are different.
First, these models are less spherical than their warmer counterparts.
Second, the transition to non-spherical systems occurs at a smaller
value of $\murz \approx 0.40$.  Third, all systems with $\murz \ga
0.40$ form nearly prolate systems; there is no strong division between
prolate spheroidal and triaxial systems.  Overall, the distinction
between models with small and large \murz\ is diminished compared to
the warm models.  This is unsurprising since particles in these models
have small velocities, and as a result, the difference between
isotropic and radially anisotropic systems is relatively minor.  The
evolution of these cold systems is dominated by collapse.

We have evolved several of these models for much longer times to
determine the axis ratio behavior on two-body relaxation timescales.
While some of our models reach and maintain their minimum \ba, \ca\
values for up to tens of crossing times, all of our models appear to
become more spherical over a relaxation time.  This is consistent with
the behavior of models presented in \citet{ts99}.  Those authors
speculate that the long-term transformation to sphericity stems from a
violent relaxation process that follows the ROI and a subsequent
two-body relaxation that occurs between the centrally collapsed core
(treated as a single very massive particle) and particles near the
edge of the system.  While our models do not show a clear distinction
between the ROI and violent relaxation phases, there is some evidence
for a weakening of the instability as violent relaxation comes to an
end.

From the correspondence between $A_0$ and \murz, we confirm the
speculation in \citet{bgh86} that some of the variation in the value
of $A_0$ that separates stable and unstable models is due simply to
different mass distributions.  In our models, non-sphericity sets in
when: $A_0\ga 2.1$ for initially Gaussian models, $A_0\ga 2.5$ for
initially cuspy models, and $A_0\ga 3.0$ for initially constant
density models.  We interpret this trend as follows.  Models with
different density profiles have different numbers of particles in
their central regions.  Any bar-like structure that forms near the
center of a system through the ROI can grow only by ensnaring
neighboring orbits.  In less dense systems, there are fewer particles
in the central regions, and the bar-like structure will thrive only if
particles from the outer regions approach the center.  This occurs
more frequently when more of the orbits in the outer regions are
radially anisotropic, or equivalently have smaller \murz (larger
$A_0$) values.  Overall, our results support those of \citet{bgh86}
and \citet{ma85}; larger initial radial velocity anisotropy near the
center of a system makes it more susceptible to the ROI.

\subsection{Cold, Initially Isotropic Systems}\label{isomods}

We now turn to the other class of initial conditions that lead to the
ROI; systems with isotropic velocity distributions but extremely small
virial ratios, $\qz \la 0.1$.  In particular, we are interested in the
behavior of \mur\ during the evolution of systems in the small-$\qz$
cases.  Initially, $\murz=0$, but \mur\ later increases dramatically
as any very cold collapse leads to `shell-crossing'.  Sets of
particles with similar initial radial positions fall to the center of
the potential well and appear to rebound radially outward where they
coexist with other sets of particles that are infalling.  This overlap
of velocity behavior gives rise to large radial velocity dispersions
which, in turn, increases $\beta$ in the overlap region.

Following the work of \citet{u93} and \citet{hetal96}, we investigate
how a) the maximum \mur\ achieved by the small-$\qz$ cases links to
the triaxiality of the resulting system and b) these systems'
axis ratio behavior compares to those of initially anisotropic systems.
The hypothesis is that the maximum \mur\ will determine global
triaxiality and that the \mur--axis ratio relation observed in the
initially anisotropic systems will continue to hold.  Five initially
isotropic models with $\qz=0.00$, 0.02, 0.04, 0.06, and 0.08 have
been evolved for each of the three initial density profiles discussed
earlier (see \S~\ref{inits}).

\subsubsection{$\rho_{\rm initial}(r)=\rho_0$}

The initially constant density models undergo such strong collapses
that a sizeable fraction of particles ($\approx 35\%$) escape the
system and leave a mostly spherical core for all values of $\qz$.
Except in the coldest systems ($\qz=0.00$ and 0.02), the axis ratios
for these models typically remain $\ga 0.85$.  The coldest models
briefly becomes oblate spheroidal with $\ca \approx 0.75$.  The
maximum values of \mur\ (\murmax) for these models appear to be
approximately 0.5.  However, this is a result of coarse time
resolution of these simultions; \mur\ can reach larger values, but
only for short intervals $\la 0.1$ \tc (see \S~\ref{qltconst}).  Fully
isotropic models with slightly ``warmer'' initial conditions
($\qz=0.1, \: 0.2$) have similar maximum values of \mur, mass loss
fractions, and axis ratio evolutions.  How do these models compare to
their most similar anisotropic counterparts?  The $\qz=0.1$ models
with the weakest initial anisotropies ($\murz=0.30,0.50$) have smaller
mass loss fractions ($\approx 15-20\%$), but still result in a
spherical core.  We conclude that the ROI is suppressed in these
systems simply because sufficient mass occupying radially anisotropic
orbits is not retained during the collapse.  We point out that not all
systems that lose mass necessarily remain spherical.  If $\murz \ga
0.50$, then a system may lose mass and still have enough residual
radial anisotropy to bring on the ROI.

\subsubsection{$\rho_{\rm initial}(r) \propto r^{-1}$}

Models with this central density cusp lose approximately 15\% of their
mass after these cold collapses.  Unlike the constant density cases
above, these models all reach maximum \mur\ values $\ga 0.85$ during
their evolutions.  According to our hypothesis, these systems should
all undergo the ROI and become triaxial.  This is observed, but the
$\qz=0.08$ model remains nearly spherical for the first 4 crossing
times.  After that, the inner 80\% of the mass becomes prolate while
mass exterior remains spherical.  This change does not coincide with
the mass loss which occurs after $\approx 8$ crossing times.  For
models with $\qz<0.08$, the trends seen in initially anisotropic
systems, a) weaker anisotropy produces more prolate objects and b)
stronger anisotropy produces smaller \ca\ values, are also reproduced.
The increased central potential due to the cusp allows the system to
better retain its mass through the collapse, which in turn allows
larger radial anisotropies to develop and lead to the ROI.

\subsubsection{$\rho_{\rm initial}(r) \propto e^{-r^2}$}

Initially Gaussian density profiles provide a centrally concentrated
system without a cusp.  With our choice of Gaussian scalelength (see
\S~\ref{inits}), the $\qz<0.1$ models lose $\la 1\%$ of their mass
after their collapses.  Like the cuspy models above, these systems all
reach maximum \mur\ values $\ga 0.80$ and do not remain spherical.
However, while the models with $\qz < 0.08$ follow the previously
discussed trends (prolate-weaker anisotropy, smaller \ca-stronger
anisotropy), the $\qz=0.08$ model remains nearly spherical throughout
its evolution ($\ba,\: \ca >0.85$).  We have tested the robustness of
this result by evolving 5 different realizations of the $\qz=0.08$
model.  The maximum variation between axis ratios in the realizations
is $\approx 0.1$; all realizations stay nearly spherical. 

\subsection{Synthesis of Anisotropic and Cold, Isotropic Global
Results}\label{globsynth}

We have found that $\qz$ can be a useful ROI discriminator for
initially isotropic systems.  \citet{ma85} conclude that $\qz \la 0.1$
should separate ROI stable and unstable models that form from
isotropic, spherical collapses.  Our boundary value appears to depend
on the initial density profile, but is $\approx 0.1$ for initially
cuspy ($\rho \propto r^{-1}$) and Gaussian profiles.  For initially
constant density systems, these cold collapses are apparently too
violent to allow the ROI to fully develop as large fractions of mass
escape.  However, the coldest systems develop non-spherical shapes
before substantial mass loss occurs and then return to nearly
spherical afterwards.  We have also evolved systems with larger $\qz$
values.  We find that during these evolutions, axis ratios typically
decrease by less than 5\% for $\qz \ge 0.1$.  For Gaussian initial
densities, the axis ratios decrease by less than 2\% for $\qz \ge 0.1$.
Hence, there is a dramatic difference in evolution between systems
with $\qz \ge 0.1$ and $\qz < 0.1$.  In this class of models, the
global maximum value of \mur\ does not change significantly with
$\qz$, but there is a trend for colder collapses to result in larger
\murmax.  This is not surprising and we conclude that the value of
\murmax\ is not a useful diagnostic for the ROI.

$\qz$ also plays a role in the behavior of the ROI for initially
anisotropic systems, along with \murz.  Systems with $\qz > 0.1$ reach
similar levels of non-sphericity when they have similar amounts of
mass on radially anisotropic orbits, independent of initial density
profile.  In general, when $\murz \la 0.60$, systems remain spherical.
Higher values of \murz\ lead to evolutions that produce prolate and
triaxial systems.  As with the initially isotropic cases, $\qz \approx
0.1$ is again a boundary for ROI behavior.  Initially anisotropic
models with $\qz =0.1$ have substantially lower thresholds to achieve
non-sphericity; $\murz \ga 0.40$ for cuspy and Gaussian models.
Constant density models remain roughly spherical for $\murz \la 0.60$
and become triaxial with more radial velocity distributions.

\begin{figure}
\plotone{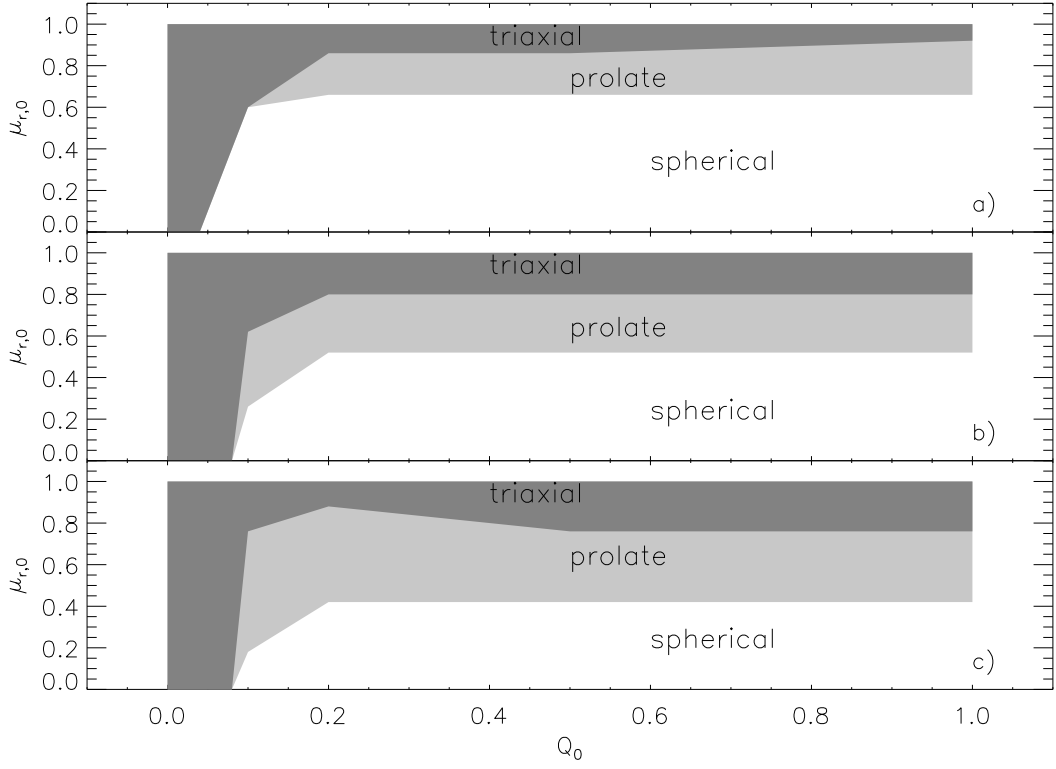}
\figcaption{A visual summary of the global behavior of the ROI for
initially isotropic and anisotropic models.  Panels a, b, and c
correspond to models with constant, cuspy, and Gaussian initial
density profiles, respectively.  Models with initial conditions in the
dark gray areas will evolve to triaxial shapes at the peak of the ROI.
Initial conditions in the light gray region result in prolate systems.
Systems that start from the unshaded region will remain more-or-less
spherical throughout their evolutions.  For the constant models in
panel a, note that the models with $\qz \la 0.1$ had significant
amounts of mass loss.  The shape of the final virialized system refers
only to the remnant, and drawing strong conclusions about the ROI from
these cold models is problematic.
\label{area}}
\end{figure}

A visual summary of these results is presented in Figure~\ref{area}.
Panels a, b, and c correspond to models with constant, cuspy, and
Gaussian initial density profiles, respectively.  The shaded areas
represent the initial conditions that give rise to various system
shapes at the peak of the ROI.  For small \qz, every model
investigated became non-spherical.

\section{Onset and Early Development of the ROI}\label{roidev}

We now compare the development of the ROI in isotropic and anisotropic
systems, describe how the onset of the ROI depends on $\qz$, and
investigate how the velocity distributions in these systems change
during evolution.  Section~\ref{roiea} is intended to complement the
discussion in \S~\ref{anisomods}, while Sections~\ref{roiei} and
\ref{isomods} are counterparts.

\subsection{Initially Anisotropic Systems}\label{roiea}

To complement our discussion of the global behavior of the ROI for
initially anisotropic systems and compare to isotropic cases, we now
turn to describing the onset of the instability in anisotropic models.
We have run high time resolution versions of a subset of the
anisotropic models discussed earlier.  These evolutions cover
approximately the first two initial crossing times for each model.  We
again focus on models with $\qz$-values of 1.0, 0.5, 0.2, and 0.1.  To
cover the same range of initial anisotropy as discussed in
\S~\ref{anisomods} , models for each $\qz$ were evolved for $r_a=0.9$
(most isotropic), 0.5, and 0.1 (most radially anisotropic).
Variations in initial density profiles produce qualitatively similar
results, which we discuss first.  Later we compare the behavior of our
models to the analytical predictions from \citet{hetal96}.

\subsubsection{Mass Fraction \& Axisratio Behaviors}

Models with $r_a=0.9$ have the smallest amounts of initial radial
anisotropy for each density profile.  Initially constant density
models have $\murz=0.50$, cuspy models have $\murz=0.40$, and models with
initially Gaussian profiles have $\murz=0.10$.  For all values of
$\qz$, the value of \mur\ rapidly decreases up to the time of maximum
collapse (when the system is in its most compact configuration), with
\mut\ increasing to a maximum during the same period.  After maximum
collapse, \mur\ values rise again with colder models showing stronger
and more rapid increases.  In the coldest systems ($\qz \le 0.2$),
\mur\ can reach and surpass the threshold values discussed in
\S~\ref{anisomods}.  However, the value of the centrally-concentrated,
isotropic mass fraction \muic\ remains greater than about 0.20.  During
all of these evolutions, the axis ratios maintain quite spherical
values ($\ba \ga 0.90$ and $\ca \ga 0.85$).  Deviations from
sphericity correspond to times during which \mur\ is near or above its
threshold value and \muic\ is near its minimum value.

More initial radial anisotropy occurs in systems with $r_a=0.5$;
\murz=0.90, 0.80, and 0.60 for models with initially constant, cuspy,
and Gaussian density profiles, respectively.  As with the more
isotropic models, the \mur\ values drop to their minimum levels just
before maximum collapse in these evolutions.  The \mui\ and \mut\
values both increase during this time, with warmer models showing
larger \mut\ levels and colder models having larger \mui\ values.
After maximum collapse, \mur\ values increase to at least threshold
levels for all \qz\ values.  In the warmer models, \muic\ values
remain larger than 0.20 during the interval when \mur\ is above the
threshold.  Axisratios for these warm models do not change
significantly from spherical values ($\approx 0.90$) during this part
of the evolution.  Colder models have $\muic \la 0.20$ while \mur\ is
above the threshold, and their axis ratios decrease while both of these
conditions hold.  Overall, the amount of axis ratio change increases
with decreasing \qz\ even as the time at which the axis ratios reach
their minimum values decreases.  Minimum axis ratios are typically
reached after several maximum collapse times have passed.  Constant
density models find their axis ratio minima earlier than cuspy models.
Axisratios in Gaussian density models take the longest amount of time
to reach their minimum values, around 20 maximum collapse times.

When $r_a=0.1$, the models are initially completely radially
anisotropic for all density profiles.  Since no mass is isotropic, our
previous findings would suggest that these systems should immediately
begin to change shape.  However, no models show substantial changes in
axis ratios until after their times of maximum collapse.  Minimum
axis ratio values are typically reached between 4 and 8 maximum
collapse times, independent of initial density profile.  In these
extreme models, the time evolution of \mur\ for initially constant and
cuspy systems behaves somewhat differently than for Gaussian systems.
In constant and cuspy models, \mur\ decreases to at least threshold
values by the time of maximum collapse and then increases again.  The
\mur\ values in the Gaussian models remain above threshold values for
at least the first two crossing times.  The outer edges of constant
and cuspy models develop significant levels of isotropy ($\mui \ga
0.40$), while Gaussian models do not show this behavior.  Despite these
differences, all models share the following behavior; the amount of
centrally-concentrated, isotropic orbits increases during an
evolution, and the increase is greater for colder initial conditions.
In general, \muic\ reaches a value of 0.20 between 1.5 and 2.0 \tc.
Again, as discussed previously, the strength of the axis ratio change
in a given system increases with decreasing \qz.

\subsubsection{Velocity Dispersion Evolutions}

\citet{hetal96} present predictions [based on work in \citet{ketal96}]
for the evolution of velocity dispersions in systems with initially i)
isotropic velocity distributions and ii) constant and cuspy ($\rho
\propto r^{-1}$) density profiles as well as results of numerical
models.  We directly compare our results to theirs for analogous
models and note resemblances for our Gaussian models, which do not
have analytical counterparts in their work.

\begin{figure}
\plotone{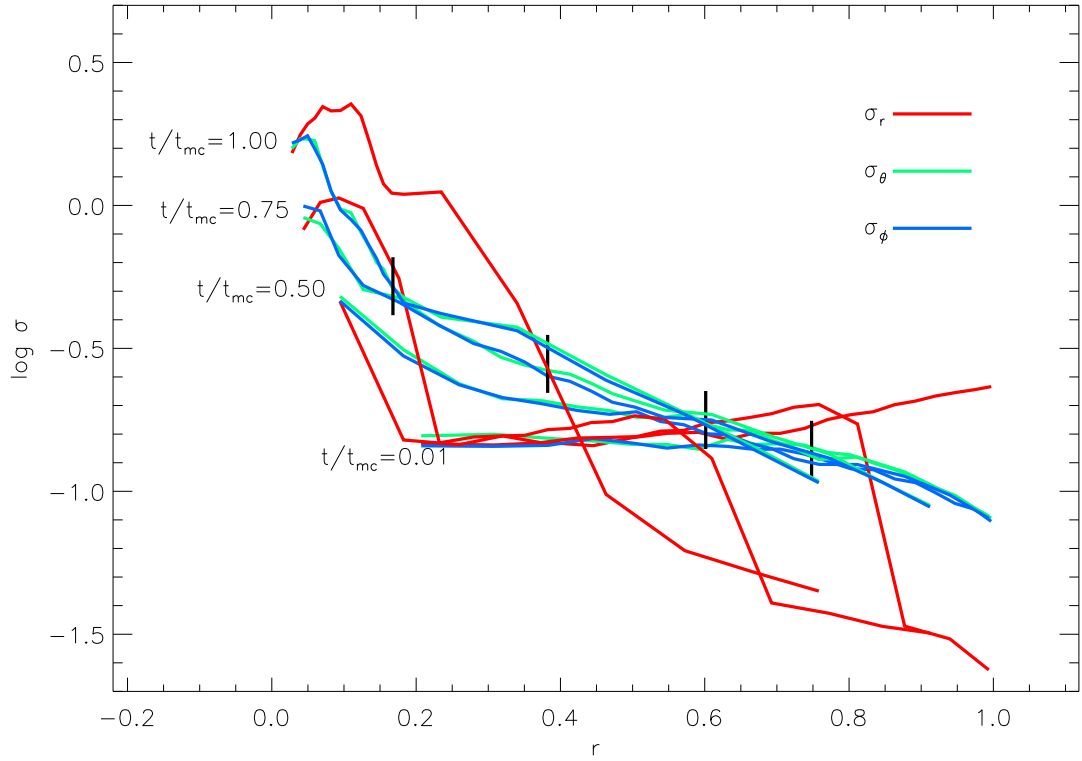}
\figcaption{Comparisons between spherical velocity component
dispersions in a cuspy, initially radially anisotropic model with
$r_a=0.9$ and $\qz=0.1$.  The overall behavior of the curves matches
Figure 5b in \citet{hetal96}.  The curves represent dispersions at
four different times during the evolution up to the time of maximum
collapse $t_{\rm mc}$.  The earliest set of curves is essentially the
beginning of the simulation, one-hundredth of the time to maximum
collapse.  The next set is halfway to collapse $t=0.5 \: t_{\rm mc}$,
the third set is for $t=0.75 \: t_{\rm mc}$, and the upper-most curves
show the situation at the time of maximum collapse.  The vertical
lines show the location of the half-mass radius at each interval.
Note that the half-mass radius decreases as the time of maximum
collapse is approached.  With this level of initial anisotropy, the
axis ratios do not reach their minimum values until many maximum
collapse times have passed.
\label{disp1}}
\end{figure}

Tangential dispersions ($\sigma_{\rm tan}$) in our initially constant
density and cuspy $r_a=0.9$ models \citep[which are most similar to
the completely isotropic models in][]{hetal96} evolve at least as fast
as radial dispersions ($\sigma_r$), in qualitative agreement with
analytical models.  In particular, colder ($\qz=0.1$ and 0.2) models
have dispersion evolutions similar to their numerical counterparts in
\citet{hetal96}.  Our Gaussian models have dispersion evolutions that
are qualitatively similar to those for the cuspy models.  For example,
Figure~\ref{disp1} shows how the dispersions in a cold ($\qz=0.1$),
cuspy model evolve up to the time of maximum collapse and is quite
similar to Figure 5b in \citet{hetal96}.  We note that our models
disagree with the analytical predictions in much the same way that the
numerical models in \citet{hetal96} do.  The lack of inifinite
contraction in numerical models forces them to behave as though they
have roughly constant density cores, which analytical models predict
should evolve isothermally.  The analytical models also neglect system
edges, which leads to numerical models with smaller than predicted
dispersions in their outer regions.  While we find good agreement with
the previous numerical models, there are two aspects of our evolutions
that are different.

First, our warm models ($\qz=1.0$ and 0.5) do not maintain small
isotropic cores as maximum collapse is achieved.  Instead, these
systems become almost completely tangentially anisotropic by the time
of maximum collapse.  \citet{hetal96} only considered models with $\qz
\la 0.3$, so this behavior has not been described before.  Second, for
all our models, the decrease in radial dispersions in the outer
regions is more severe, in magnitude and abruptness, than the previous
models.  Rather than a smooth transition between inner regions where
$\sigma_r$ increases to outer regions where $\sigma_r$ decreases, we
see a ``break'' radius that divides the inner and outer regions.  This
break radius moves inwards as the time of maximum collapse is
approached.

The break radius arises from the initial conditions we have set in our
models.  Radial dispersions are created by giving particles both
positive and negative radial velocities.  Particles near the edges of
systems have the largest anisotropies, and so, radial velocities as
well.  Particles that have positive radial velocities quickly expand
the system.  When mass shells are then created, the outermost shells
are comprised entirely of expanding particles; the average velocity is
large and positive, but the dispersion is nearly zero.  As time
passes, particles with positive radial velocities from inner regions
expand to a point where they are surrounded only by particles with
similar motions, decreasing the dispersion.  In this way, the outer
low-radial dispersion region grows inward with time.  By the maximum
collapse time, the break radius is replaced by a smooth transition
between the inner, $\sigma_r$- increasing, regions and outer,
$\sigma_r$-decreasing, regions.

Models with more radially anisotropic velocity distributions, \ie\
those with $r_a=0.5$ and $r_a=0.1$, also behave qualitatively the same
as their more isotropic cousins.  Tangential dispersions evolve very
similarly to the predictions of analytical models; they increase
throughout the system, but the largest increase occurs at the center.
Radial dispersions increase at the centers of systems to maintain
near-isotropy and decrease near the edges.  The biggest difference
between these models and the $r_a=0.9$ versions is the magnitude of
the dispersions.  For the cuspy model with $r_a=0.1$, the initial
radial and tangential dispersion values are roughly constant in
radius, but $\sigma_r \approx 30 \sigma_{\rm tan}$.  So, even as
$\sigma_{\rm tan}$ grows, the systems can stay quite radially
anisotropic.

\subsubsection{Unified Picture of ROI Onset in Initially Ansotropic
Systems}\label{anisopic}

A common story emerges from comparing models with different initial
density profiles and initially anisotropic velocity distributions.
The amount of centrally-concentrated mass on isotropic orbits plays an
important role, as previously hypothesized in \citet{ma85}.  However,
the \muic\ value, by itself, does not correlate with a system
undergoing the ROI.  We have found that a combination of \mur\ and
\muic\ can explain our simulations in the following way.  If $\muic
\la 0.20$ and \mur\ reaches levels larger than the appropriate
threshold value, then the axis ratios decrease rapidly.  We stress
that, individually, the two conditions do not describe the onset of
the ROI.  Larger values of \mur\ give rise to larger changes in
axis ratios, as long as \muic\ is small enough.  The $r_a=0.1$ models
give an interesting caveat to our picture.  These models begin totally
radially anisotropic, but the substantial changes in axis ratios do not
begin until around the time of maximum collapse.

Keeping with the picture of the radial orbit instability suggested by
\citet{pp87}, any non-spherical distribution of mass creates torques
on orbits not aligned with the major axis of the system.  We note that
this idea also has precedents in \citet{lb79} and \citet{fp84}.  These
torques would be largest when the system is most compact.  In
Appendix~\ref{torque}, we discuss a model of this situation; a test
mass and two off-center masses.  This toy model results in a torque on
the test mass that increases as the test mass radius decreases, which
we contend will hold for more realistic bars as well.  As a result of
torque increasing during the collapse, more orbits will be trapped by
the bar and further strengthen the bar.  

This idea helps explain why the warm $r_a=0.1$ models take longer to
form weaker bars than the cold models with the same radial anisotropy.
In the warm models, only a fraction of the mass will converge to the
center, so the torques that result are smaller and require a longer
time to have an effect.  Colder models allow more mass to collapse,
providing a stronger bar and torques.  We note that this picture also
qualitatively explains why more isotropic models form weaker bars.
When the orbits of particles are not nearly radial, and have
appreciable angular momentum, the torques exerted cannot align those
orbits to the bar as quickly.  

The question becomes, why don't these systems become extreme bars?
Our picture supports the hypothesis forwarded by \citet{aetal07},
namely that an instability of individual orbits along principal axes
in triaxial systems isotropizes the orbits and halts bar growth.  In
looking at the anisotropic mass fractions, we have pointed out the
important role of the centrally-concentrated isotropic mass, \muic.
The value of this quantity is closely linked to the growth of a bar.
In cases where \muic\ is sufficiently low and \mur\ is high, bar
growth occurs, followed by an increase in \muic.  We equate this
growth in \muic\ with the outcome of the orbit instabilities present
in the triaxial system.

\subsection{Cold, Initially Isotropic Systems}\label{roiei}

The hypothesis [also discussed by \citet{hetal96}] is that \mur\ for
an initially isotropic system reaches the level indicated by the
initial anisotropic models at the same time the system begins to
deviate from spherical symmetry.  In effect, the isotropic system
develops radial anisotropy through collapse beyond which point it
behaves like an initially anisotropic system.  To test this idea, we
look at high time resolution versions of the cold, isotropic models
discussed in \S~\ref{isomods}.

Like \citet{hetal96}, we find that initially isotropic systems with
$\qz < 0.1$ remain nearly, but not perfectly, spherical during their
collapse phases.  Specifically, the axis ratios slowly decline from
their initial values until just before the point of maximum collapse,
when the half-mass radius is minimum.  We attribute this slow initial
change in axis ratio values to the type of evolution discussed in
\citet{lms65}.  Any slight departure from sphericity becomes magnified
during gravitational collapse.  The axis ratio change is expected to
increase as the rate of collapse increases, or alternatively, as the
system gets colder.  This is indeed what is seen for all of our
models, independent of density profile, but the effect is more
prominent in the initially constant density models.  As an example of
this trend from the constant dentisy models, the $\qz=0.08$ model has
$\Delta \ca = -0.04$ during the time until maximum collapse, while the
$\qz=0.00$ model has $\Delta \ca = -0.18$ over the same amount of
time.  The \ca\ values given here refer to the axis ratios of the
innermost 95\% of the mass.  The 80\%-mass axis ratios show similar,
but more exaggerated, behavior since their initial axis ratio values
tend to be initially farther from 1.0, compared to their 95\%-mass
counterparts.  We now continue with more detailed descriptions of the
early evolutions of these systems (from $t=0$ to $t=2 \tc$).

\subsubsection{$\rho_{\rm initial}(r)=\rho_0$}\label{qltconst}

In general, the constant density cases maintain a large degree of
isotropy throughout their infall periods.  We consider the infall
period to be the time from the beginning of the simulation to the
maximum collapse time, $t_{\rm mc} \approx 1.1 \tc$ for these models.
The $\qz=0.00$ and $\qz=0.02$ models maintain isotropy in at least the
inner 50\% of the mass in the system continuously until the maximum
collapse time.  For models with $0.04 \le \qz \le 0.08$, $\mui \ge
0.50$ until $\approx 0.1 \tc$ before the maximum collapse time.
Between this time and the maximum collapse time, the central isotropic
mass fraction \muic\ decreases as more and more orbits become
tangentially anisotropic.

At the maximum collapse time, all systems rapidly develop large
amounts of radial anisotropy, with a corresponding decrease in central
isotropy.  This is due to particles rebounding outward from their
collapse occupying the same volume as other particles which are still
infalling.  This period of radially anisotropic domination is very
short-lived.  The duration is about $0.1 \tc$ for the $\qz=0.08$
model, and decreases with $\qz$.  After the brief period of radial
anisotropy, \muic\ increases to its pre-maximum collapse value.

When \muic\ briefly decreases, the axis ratios change abruptly (for
both the 80\%-mass and 95\%-mass versions).  These changes are not
monotonic; the axis ratios can increase and decrease by sizeable
amounts, but by the time \mur\ reaches its maximum, the axis ratios
have decreased to values lower than they were at maximum collapse
time.  This overall decrease scales inversely with the $\qz$-value.
For the coldest models, axis ratios reach their minimum values soon
after the maximum collapse time.  Once \muic\ returns to its
pre-maximum collapse value, the axis ratios do not change appreciably
until mass begins to escape from the system.  This occurs fairly
rapidly for the $\qz=0.00$ model (mass loss begins at $\approx 1.6
\tc$), but takes longer to develop (after $2 \tc$) for models with
larger $\qz$.

\begin{figure}
\plotone{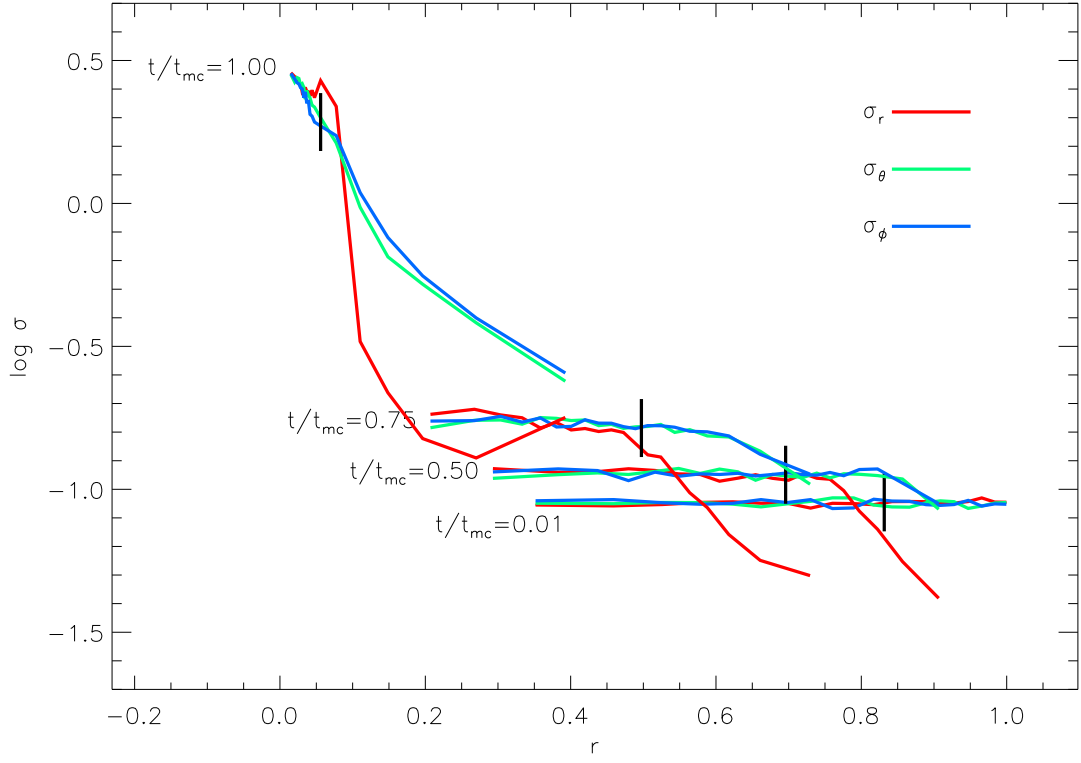}
\figcaption{Comparisons between spherical velocity component
dispersions in the initially isotropic model with $\qz=0.04$.  The
overall behavior of the curves matches Figure 3b in \citet{hetal96}.
The curves shown have the same meaning as those in Figure~\ref{disp1}.
Again, take note that the half-mass radius marks move to the left in
the figure as time increases.  Axisratios typically reach their
minimum values after roughly two---three maximum collapse time for
these models.
\label{disp2}}
\end{figure}

We find that the velocity dispersions in the models with $\qz \ge
0.02$ evolve similarly to the constant density model discussed in
\citet[][compare our Figure~\ref{disp2} with their Figure
3a]{hetal96}.  Initially, both the radial and tangential dispersions
are equal and constant throughout the system.  As the system evolves
to about 2/3 of the maximum collapse time, the dispersions in the
central regions grow contemporaneously, while the radial dispersion
grows much more slowly than the tangential dispersions beyond about
the half-mass radius.  During the final third of the evolution to
maximum collapse, the innermost dispersions grow much more rapidly
than those farther out in the system.  The central half of the system
is isotropic while the outer half is very tangentially anisotropic.
The $\qz=0.00$ model on the other hand shows a different behavior,
much more like the analytical results in \citet {hetal96} [based on
work in \citet{ketal96}].  The radial and tangential dispersions
evolve together for approximately the first 3/4 of the maximum
collapse time, maintaining system-wide isotropy.  As the system
continues towards maximum collapse, the radial dispersion begins to
lag behind the tangential dispersions; approximately 30\% of the
outermost mass is on tangentially anisotropic orbits.

\subsubsection{$\rho_{\rm initial}(r) \propto r^{-1}$}\label{qltcusp}

These models maintain central isotropic regions for approximately $0.5
\tc$.  The maximum collapse times for these models are $t_{\rm mc}
\approx 1.1 \tc$.  For the $\qz=0.00$ case, the decrease in central
isotropy after $0.5 \tc$ is accompanied by a growth in radial
anisotropy with very little tangential anisotropy present.  Models
with $\qz \ge 0.02$ also lose their central, isotropic regions, but
they develop a substantial amount of mass on tangentially anistropic
orbits ($\mut = 0.8$).  The loss of central isotropy persists for $0.5
\tc \la t \la 1.5 \tc$ for the $\qz=0.00$ and $\qz=0.02$ models.  As
$\qz$ increases, it takes longer for the central isotropy to return to
pre-maximum collapse levels.  In the $\qz=0.08$ model, \muic\ drops to
and remains around 0.15 until about 4 \tc.  For those models that
develop large amounts of tangential anisotropy, it disappears prior to
maximum collapse.  As the tangentially anisotropic mass fraction
decreases, \mur increases, typically reaching its maximum value
shortly after maximum collapse.

The axis ratios (both 80\%-mass and 95\%-mass) for these systems show
slow decreases (like those expected simply from contraction) until the
\mur values reach the threshold level discussed in \S~\ref{anisomods}.
At this point, \muic\ is at its minimum and the axis ratios decline
more sharply.  This decline continues until $\muic \ga 0.25$.  As
discussed earlier, this minimum in the central isotropic mass fraction
lasts for shorter periods of time for models with smaller $\qz$.  At
the same time, the total axis ratio change is inversely proportional to
$\qz$, so the axis ratio decrease is most rapid for lower-$\qz$
systems.  Overall, the axis ratios in the coldest systems reach their
minimum values earlier ($\approx 2$ \tc) than those in warmer systems
(after 4 \tc).

A longer simulation of the $\qz=0.08$ model with the same time
resolution gives some further insight.  The 95\%-mass axis ratios
remain at nearly spherical values for the first 8 crossing times
(until mass loss occurs).  The 80\%-mass axis ratios remain nearly
spherical until $\approx 4 \tc$, after which they decline together
indicating a moderately strong bar has formed.  The axis ratios (both
95\%-mass and 80\%-mass) for the initially isotropic $\qz=0.1$ model
remain above 0.9 until a mass loss event occurs.  We have evolved an
additional cuspy model with $\qz=0.09$ to better pin down the boundary
of systems that are susceptible to the ROI.  The axis ratios for the
$\qz=0.09$ model behave similarly to the $\qz=0.08$ case.  We
interpret this behavior as evidence that $\qz=0.1$ is very nearly the
value that separates ROI stable and unstable systems for this density
profile.  When $\qz$ is near 0.00, the instability occurs
simultaneously throughout a system.  Systems with borderline-unstable
$\qz$ values show delayed onset of the instability and once the
instability begins, only the inner part of the system appears to take
part.

As with the constant density models, the evolution of the velocity
dispersions in the cuspy systems with $\qz \ge 0.02$ also agree
with the behavior of similar systems discussed in \citet{hetal96}.
Until about half of the maximum collapse time, the systems roughly
follow this behavior: dispersions in the central regions grow rapidly
but isotropically, the tangential dispersions increase more slowly
near the edges, and the radial dispersions remain nearly constant
outside the central $\approx$ 10\% of the mass.  The $\qz=0.00$
model differs during these early stages of the collapse in that the
dispersions evolve isotropically throughout the system.  As the
systems continue to maximum collapse, radial anisotropy begins to
dominate from the center outward (independent of $\qz$).  By
three-quarters of the maximum collapse time, the central 30-50\% of
the mass is on radially anisotropic orbits.  At the maximum collapse
time, the central-most 5-10\% of the mass is isotropic, with the
remainder being radially anisotropic.  \citet{hetal96} explain the
isotropic evolution of the central dispersions by stating that the
density remains nearly constant (spatially) in these regions; isotropy
is then the expected behavior (see \S~\ref{qltconst}).  We find that
the density profile slowly changes from the initial $r^{-1}$ shape to
a near-de Vaucouleurs shape at the maximum collapse time, maintaining
a cuspy center (even if it is not exactly $r^{-1}$).

\subsubsection{$\rho_{\rm initial}(r) \propto e^{-r^2}$}

Compared to the previous systems, these models have the shortest
maximum collapse times, $t_{\rm mc} \approx 0.5 \tc$.  The central
isotropic mass fraction begins to decrease about 0.1 \tc\ prior to the
maximum collapse.  Like the situation for cuspy profiles, the
$\qz=0.00$ model mainly develops strong radially anisotropic motions
as the isotropy is lost; only a small fraction of mass ($\approx
10\%$) is tangentially anisotropic.  Models with $\qz \ge 0.02$ lose
central isotropy as well, but they rapidly become near-completely
tangentially anisotropic before maximum collapse.  This tangential
anisotropy is quickly lost as the system develops large radially
anisotropic mass fractions ($\approx$ 0.50) by the maximum collapse
time.  The \mur\ values reach maximum values of around 0.80,
independent of $\qz$, approximately 0.1 \tc\ after maximum collapse.
The central isotropic mass fraction dips to 0.05 before maximum
collapse, typically when most of the mass is on tangentially
anisotropic orbits.  For the $\qz=0.00$ case, the value remains at
this minimum value during the entire rise of \mur.  In models with
$\qz \ge 0.02$, \muic\ increases slightly by the time of maximum
collapse.  The amount of increase is proportional to $\qz$; for
example, \muic\ increases to 0.10 for $\qz=0.02$ while \muic\
increases to 0.20 for $\qz=0.08$.

The behavior of axis ratios in these models is similar to those for the
cuspy models.  The models present slow initial declines in axis ratios
until the radially anisotropic mass fraction increases above the
threshold value for this density profile ($\approx 0.5$), occurring at
very nearly the time of maximum collapse.  At this point in time, the
axis ratios start to decline more rapidly.  This rapid decline
continues until \mur\ reaches its maximum and while \muic\ is less
than $\approx 0.20$.  As before, the total decrease in axis ratio
values is inversely proportional to $\qz$.  Since all models reach
\murmax\ at approximately the same time, colder initial conditions
produce steeper axis ratio changes.  Like the cuspy models, we again
see a trend for the axis ratios in the coldest models to reach minimum
values earlier (between 1 and 2 \tc) than warm models (after 3 \tc).

Velocity dispersions for these models evolve differently than in
either the constant or cuspy density profile cases.   The
$\qz=0.00$ model dispersions stay isotropic for roughly half the
time to maximum collapse.  By 75\% of the maximum collapse time, the
radial dispersion has a) evolved with the tangential dispersions at
the very center, b) lagged behind tangential values out to the
half-mass radius, and c) reached values larger than tangential beyond
the half-mass radius.  At the maximum collapse time, the system is
isotropic at the very center and around a radius about twice the
half-mass radius, but is radially anisotropic everywhere else.  Models
with $\qz \ge 0.02$ have different dispersion behaviors.  The
radial dispersions actually decrease in time over most of the radial
extent of the systems.  Since the tangential dispersions increase
during the same time, these systems develop large amounts tangential
anisotropy.  The amount of radial dispersion decrease is linked to
$\qz$; lower $\qz$ leads to smaller decreases.  The very centers
of all of these models remain isotropic throughout the maximum
collapse time.  By the maximum collapse time, the radial dispersions
have grown faster than the tangential dispersions inside the half-mass
radius but continue to lag them outside this radius.

\subsubsection{Unified Picture of ROI Onset in Cold, Initially
Isotropic Systems}\label{isopic}

As previously discussed for initially anisotropic systems
(\S~\ref{anisopic}), the roles of \mur\ and \muic\ are key to
understanding if and how the radial orbit instability will occur.
If \mur\ rises above the appropriate threshold value and \muic\ is
small ($\la 0.20$), then a system will change its shape dramatically.
The exact amount of axis ratio decrease depends on the difference $\mur
-\muic$ during this interval.  Colder initial conditions lead to
smaller minimum \muic\ values, and larger axis ratio changes.  Then,
independent of the \mur\ value, when \muic\ increases above $\approx
0.20$, the axis ratios do not substantially change further.

\begin{figure}
\plotone{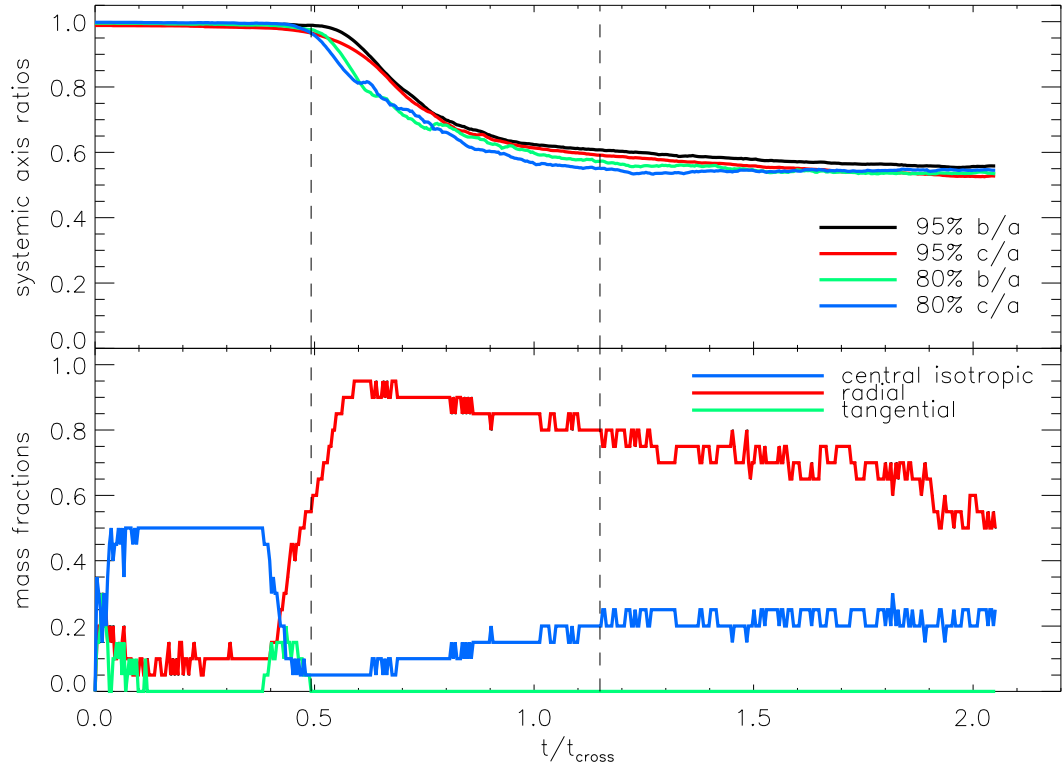}
\figcaption{The top panel shows the axis ratios as a function of time
in the initially isotropic, Gaussian density profile system with
$Q=0.00$.  The behavior of the velocity mass fractions is plotted in
the bottom panel.  The vertical dashed lines mark the times at which
a) left line, 55\% of the system mass is on radially anisotropic
orbits and b) right line, 20\% of the central mass follows isotropic
orbits.
\label{isolink}}
\end{figure}

As discussed in \S~\ref{anisomods}, the threshold values of \mur\ for
systems with $0.2 \le \qz \le 1.0$ are: $\approx 0.65$ for constant
density, $\approx 0.60$ for $\rho \propto r^{-1}$, and $\approx 0.55$
for Gaussian initial density.  The initially isotropic systems develop
these \mur\ values at approximately the same times that the axis ratios
begin to decrease in every case.  For example, the Gaussian density
system with $\qz=0.00$ has a well-defined axis ratio drop that starts
just before $t=0.5 \tc$.  In this model, $\mur=0.55$ when $t=0.48
\tc$.  The axis ratio and anisotropic mass fraction behaviors early in
the evolution ($t< 2 \tc$) are shown in Figure~\ref{isolink}.  The
vertical dashed lines indicate the times at which $\mur=0.55$ (left)
and $\muic=0.20$ (right).  Note the changes in the axis ratio behaviors
at these times.  In general, the axis ratios for cold, initially
isotropic models reach their minimum values earlier (after
approximately 2 or 3 maximum collapse times) than those for
initially anisotropic models (after approximately 5--10 maximum
collapse times).

\section{Summary \& Conclusions}\label{summary}

We have created a variety of self-gravitating, $N$-body systems to
investigate the behavior of the radial orbit instability (ROI).  Our
models use modest particle numbers ($N=10^4$, $10^5$), but have been
shown to simulate collisionless behavior for at least one two-body
relaxation time.  Tests with various softening length/particle number
combinations lead us to the conclusion that minor variations in these
quantities do not play significant roles in determining the evolution
of our models.  With fixed $N$ and softening length, we have created a
suite of models with various a) initial virial ratios $\qz$ (relating
the total kinetic and potential energies), b) levels of initial
velocity anisotropy, and c) initial density profiles (constant, cuspy,
and Gaussian).

In a global sense, the behavior of the ROI, in terms of its strength
and overall effect on system shape, is closely linked to the fraction
of system mass that follows radially anisotropic orbits, \mur.  We
find very clear threshold values of \murz\ in our simulations of
initially anisotropic systems that divides models into spherical and
non-spherical cases.  The exact threshold value of \murz\ depends on
the initial density profile of the model, but when $\murz \la 0.50$,
the warm and hot systems investigated remain nearly spherical
throughout their evolutions (at least 20 crossing times).  With a
modest amount of additional radial anisotropy, our models transform
from spherical shapes to very prolate, bar-like shapes.  Higher radial
anisotropy leads to systems that take on triaxial, bar-like forms.  We
find that the \murz\ quantity is a reliable predictor of the final
shape of the system, as long as the system is sufficiently warm ($\qz
\ga 0.2$).

In terms of initial anisotropy, dynamically cold systems behave quite
similarly to one another.  This is unsurprising as the systems' lack
of substantial kinetic energy necessarily relegates any differences in
velocity distributions to play a minor role.  In other words, when
particles can only have slightly non-zero speeds, it doesn't matter
how those velocities are oriented.  The early evolution of such
systems will be dominated by gravity, and the dominant motion of all
particles will be infall.  Figure~\ref{area} represents a summary of
our findings regarding the global behavior of the ROI, for both
initially isotropic and anisotropic models.

Looking at cold, initially isotropic systems, we have found that the
ROI sets in when a) the amount of mass on radially anisotropic orbits
is larger than the threshold value and b) there is little mass on
isotropic orbits at the centers of the systems.  In these models, any
radial anisotropy present must develop from free-fall motions.  During
initial collapse the mass acquires radially inward velocities.
However, since the particles fall together, no large dispersions
develop.  Only after some of the mass has ``rebounded'' away from the
center do systems develop large amounts of radial anisotropy.
Accordingly, the point of maximum radial anisotropy occurs after the
time of maximum collapse of the system.

The onset of the ROI in systems with initially radial anisotropic
distributions is also tied to the amount of radially anisotropic and
centrally-concentrated isotropic mass.  We again see that the ROI
begins when \mur\ is larger than the threshold value and \muic\ is
less than about 0.20.  However, even in systems that are completely
radially anisotropic to begin with, the ROI will not begin until after
the system has maximally ``collapsed''.  This is not a true collapse,
like in the cold isotropic systems, but refers instead to a measure of
the average radius of the system reaching its minimum value.
Typically, some fraction of the mass in the system is actually
expanding at this time.

This behavior leads us to suggest that collapses, the shrinking of at
least some part of a system, play a key role in the development of the
ROI.  When the systems are most compact, torques caused by
bar-like mass distributions are strongest.  Keeping with the idea put
forward in \citet{pp87}, these stronger torques can align more orbits
with the bar more quickly than if the system were more extended.  The
collapses allow the torquing mechanism of the ROI to act more
efficiently.

Our models also serve as evidence that the isotropy-creating orbit
instability described by \citet{aetal07} is involved as the ROI is
halted.  Note that this isotropy-creating instability is completely
different from the ROI.  We do not find extreme bars among our models;
the smallest \ca\ value is $\approx 0.4$.  So, something is halting
the torquing mechanism discussed earlier.  Certainly, a system
expansion would weaken the torques produced, just as contraction
strengthened them.  The bar would stop growing simply because it could
not continue to add more orbits to its distribution.  However, if this
scenario alone describes how the ROI stops, there would be no change
to the velocity properties of particles in the bar after it had grown;
particles that rolled along the bar previously would continue their
motions.  However, we see that the centrally-concentrated isotropic
mass fraction increases after the bar forms.  This behavior is nicely
explained by the Adams \etal\ instability.  A strong triaxial (or at
least spheroidal) mass distribution is formed.  Any orbit moving along
a principal axis is unstable to being deflected away from its current
trajectory, turning radial orbits into isotropic ones.  These
isotropic orbits will make the central regions more spherical, further
weakening the bar and the torques it can cause.

\begin{figure}
\plotone{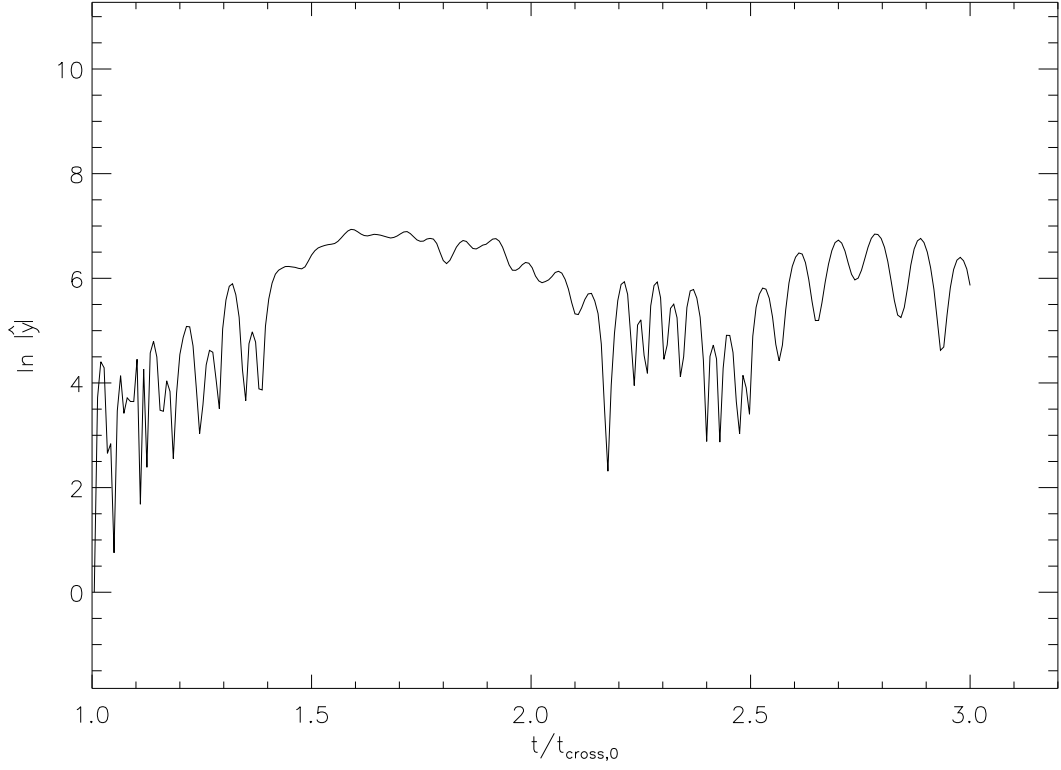}
\figcaption{The logarithm of the growth factor $\hat{y}$ for an orbit
in a high-time resolution simulation with $\qz=0.00$ and an initially
cuspy density profile.  This particle is moving in a plane
perpendicular to the intermediate axis of the bar at $t=1 \tc$ (when
the growth factor is 1).  The orbit then experiences a roughly
exponential growth perpendicular to the bar after which it settles
into a quasi-periodic oscillatory phase.  This behavior is reminiscent
of orbits that undergo the instability discussed in \citet{aetal07}.
\label{adams}}
\end{figure}

Figure~\ref{adams} illustrates one particle's motion along the
intermediate axis of the bar (formed by the ROI) as it changes after
the time of maximum collapse ($t\approx 1 \tc$).  This particle was
chosen to lie close to a plane that runs through the center of the
system perpendicular to the intermediate axis of the final bar.
Additionally, this particle has a small velocity in the intermediate
axis direction, representing our closest approximation to the orbits
studied in \citet{aetal07}.  Note that the position and velocity were
determined after the maximum collapse time.  The curve shown
represents the growth factor $\hat{y}$ described in \citet{aetal07};
the particle's intermediate axis position throughout the simulation
divided by its initial position.  Exponential growth results in linear
segments.  We see that there is a chaotic, but roughly linear, rise
starting at $t=1 \tc$ followed by a slowly oscillating but
non-increasing segment.  This behavior is similar to that shown
in Figure 4 of \citet{aetal07}.  While the likeness is rough at
best, we remind the reader that our system is still undergoing axis
ratio changes throughout the system during this interval unlike in
\citet{aetal07} where the axis ratios are spatially and temporally
fixed.  This type of behavior occurs in our simulations for orbits
with a wide range of sizes and orbit timescales.  Together with the
increasing central isotropy discussed earlier, the presence of the
Adams \etal\ instability signature in our simulations suggests that
the ROI can be halted by such an instability.  However, the number of
orbits that are vulnerable to the Adams \etal\ instability is
relatively small as particles must move nearly along principal axes.
Whether or not this population of orbits is sufficient to actually
halt the ROI must be a topic of further investigation.

\acknowledgments
The authors gratefully acknowledge support from NASA Astrophysics
Theory Program grant NNX07AG86G.  EB also thanks Shauna Sallmen for
helpful conversations regarding axis ratio determination.

\appendix

\section{Toy Model of Torques}\label{torque}

\begin{figure}
\plotone{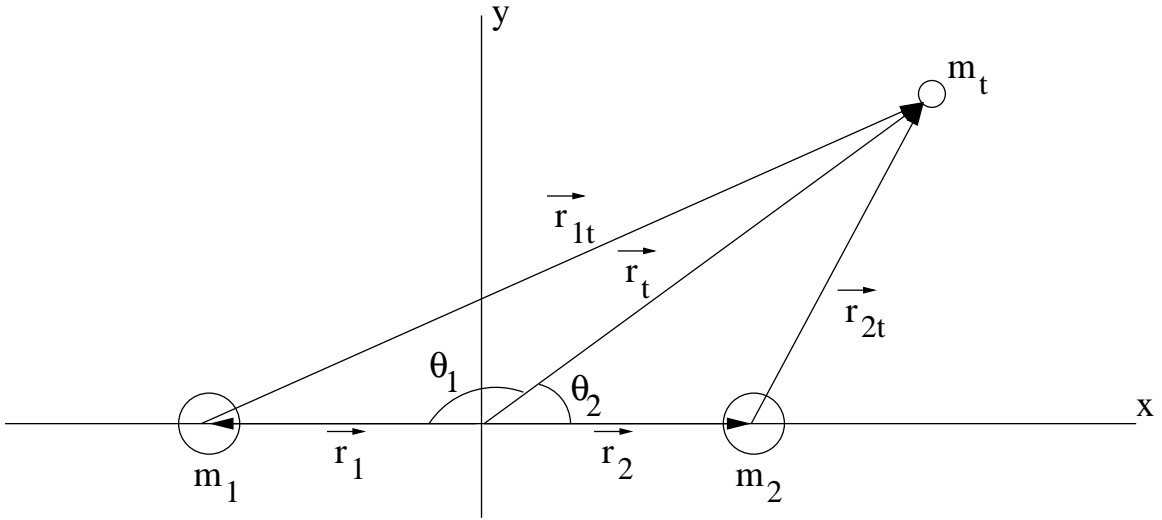}
\figcaption{The geometry of the toy model used to determine the torque
on a test mass.  The details of the model are described fully in
Appendix~\ref{torque}.
\label{torfig}}
\end{figure}

We create a simple situation to illustrate our assertion that the
torque experienced by a particle with mass much smaller than the mass
in a bar-like distribution will increase as the overall system
shrinks.  Our model consists of a test mass and two larger masses,
placed symmetrically along the $x$-axis.  Figure~\ref{torfig} shows
the geometry of the situation.  For simplicity, we consider a planar
orbit only.

The net force exerted on the test mass $m_t$ by masses $m_1$ and $m_2$
is,
\begin{equation}\label{netf1}
\vec{F}_{\rm net}=-Gm_t \left[ \frac{m_1 \vec{r}_{1t}}{r_{1t}^3} +
\frac{m_2 \vec{r}_{2t}}{r_{2t}^3} \right].
\end{equation}
Masses $m_1$ and $m_2$ are each located a distance $\ell$ from the
origin ($r_1=r_2=\ell$), so
\begin{eqnarray}
r_{1t}^2 & = & r_t^2 + \ell^2 - 2r_t \ell \cos{\theta_1} \nonumber\\
r_{2t}^2 & = & r_t^2 + \ell^2 - 2r_t \ell \cos{\theta_2}.
\end{eqnarray}
Using the fact that $\theta_1=\pi - \theta_2$, we can re-write these
distances as,
\begin{eqnarray}\label{dist1}
r_{1t}^2 & = & r_t^2\left[ 1+ \left(\frac{\ell}{r_t}\right)^2 +
2\left(\frac{\ell}{r_t}\right) \cos{\theta_2} \right] \nonumber\\
r_{2t}^2 & = & r_t^2\left[ 1+ \left(\frac{\ell}{r_t}\right)^2 -
2\left(\frac{\ell}{r_t}\right) \cos{\theta_2} \right].
\end{eqnarray}
Terms are grouped this way because we make the argument that as the
system is contracting (or expanding), the bar length should change
roughly in the same way as $r_t$.  In what follows, we assume that the
quantity $\ell/r_t$ will remain constant during collapse.

The torque about an axis through the origin (perpendicular to the
page) is just $\vec{\tau} = \vec{r}_t \times \vec{F}_{\rm net}$.  With
the net force from Equation~\ref{netf1}, we have
\begin{equation}\label{t1}
\vec{\tau}=-Gm_t \left[ \frac{m_1 \vec{r}_t 
\times \vec{r}_{1t}}{r_{1t}^3} + \frac{m_2 \vec{r}_t \times 
\vec{r}_{2t}}{r_{2t}^3} \right].
\end{equation}
We can simplify the cross-products in this expression by noting that
\begin{displaymath}
\vec{r}_{1t}=\vec{r}_t-\vec{r}_1
\end{displaymath}
and
\begin{displaymath}
\vec{r}_{2t}=\vec{r}_t-\vec{r}_2,
\end{displaymath}
so Equation~\ref{t1} simplifies to
\begin{equation}\label{t2}
\vec{\tau}=-Gm_t \left[ \frac{m_1 \vec{r}_1 
\times \vec{r}_t}{r_{1t}^3} + \frac{m_2 \vec{r}_2 \times 
\vec{r}_t}{r_{2t}^3} \right].
\end{equation}
These cross-product terms point in opposite directions; in this setup,
$\vec{r}_1 \times \vec{r}_t$ is in the $-\hat{z}$ direction and 
$\vec{r}_2 \times \vec{r}_t$ is in the $+\hat{z}$ direction.  We use
the relationship between $\theta_1$ and $\theta_2$ to write
the magnitude of the cross-products as
\begin{displaymath}
|\vec{r}_1 \times \vec{r}_t|=r_1 r_t \sin{\theta_2} = \ell r_t
\sin{\theta_2}
\end{displaymath}
and
\begin{displaymath}
|\vec{r}_2 \times \vec{r}_t|=r_2 r_t \sin{\theta_2} = \ell r_t
\sin{\theta_2}.
\end{displaymath}

Taking $m_1=m_2=m$, Equation~\ref{t2} can now be written as
\begin{equation}\label{t3}
\vec{\tau}=-Gm_t m \ell r_t \sin{\theta_2} \left[ \frac{1} 
{r_{2t}^3} - \frac{1}{r_{1t}^3} \right] \hat{z}.
\end{equation}
Using the expressions in Equation~\ref{dist1}, we reach our final
expression for the torque on the test mass in this geometry,
\begin{equation}\label{t4}
\vec{\tau}=-\frac{Gm_t m (\ell/r_t) \sin{\theta_2}}{r_t} \left[ \frac{1} 
{f_1} - \frac{1}{f_2} \right] \hat{z},
\end{equation}
where $f_1=\left[ 1 + \left(\frac{\ell}{r_t}\right)^2 -2 \left(
\frac{\ell}{r_t}\right) \cos{\theta_2}\right]^{3/2}$ and $f_2=\left[ 1
+ \left(\frac{\ell}{r_t}\right)^2 +2 \left( \frac{\ell}{r_t}\right)
\cos{\theta_2}\right]^{3/2}$.  The terms with $\ell/r_t$ or $\theta_2$
just describe the geometry of the situation, but the lone $1/r_t$
determines how the torque on the test mass changes with radial
position.  If the test mass is near the $y$-axis ($\theta_2 \approx
\pi/2$), then the torque exerted will be small independent of $r_t$
since $f_1$ and $f_2$ will be approximately equal.  Test masses near
the $x$-axis are already in the vicinity of the bar ($\theta_2 \approx
0$), and also have very little torque exerted on them.  Masses at
intermediate angles will be most affected by the torque.  More
realistic models could be constructed by adding more pairs of masses
to the system.  Superimposing pairs would add to the complexity of the
geometry term, but each pair would still have a $1/r_t$ dependence.

\end{document}